# Machine learning with observers predicts complex spatiotemporal behavior

G. Neofotistos[a, b,1], M. Mattheakis[a], G. D. Barmparis[b], J. Hizanidis[b], G. P. Tsironis[a,b] and E. Kaxiras[a,c]

[a]School of Engineering and Applied Sciences, Harvard University, Cambridge, MA 02138, USA
[b]Department of Physics, University of Crete, Heraklion 71003, Greece
[c]Department of Physics, Harvard University, Cambridge, MA 02138, USA

[1] To whom correspondence may be addressed. Email: neofotistos@physics.uoc.gr

Keywords: prediction, machine learning, chimera state, graphene, long short-term memory, reservoir computing

July 26, 2018

**ABSTRACT**
Chimeras and branching are two archetypical complex phenomena that appear in many physical systems; because of their different intrinsic dynamics, they delineate opposite non-trivial limits in the complexity of wave motion and present severe challenges in predicting chaotic and singular behavior in extended physical systems. We report on the long-term forecasting capability of Long Short-Term Memory (LSTM) and reservoir computing (RC) recurrent neural networks, when they are applied to the spatiotemporal evolution of turbulent chimeras in simulated arrays of coupled superconducting quantum interference devices (SQUIDs) or lasers, and branching in the electronic flow of two-dimensional graphene with random potential. We propose a new method in which we assign one LSTM network to each system node except for "observer" nodes which provide continual "ground truth" measurements as input; we refer to this method as "Observer LSTM" (OLSTM). We demonstrate that even a small number of observers greatly improves the data-driven (model-free) long-term forecasting capability of the LSTM networks and provide the framework for a consistent comparison between the RC and LSTM methods. We find that RC requires smaller training datasets than OLSTMs, but the latter require fewer observers. Both methods are benchmarked against Feed-Forward neural networks (FNNs), also trained to make predictions with observers (OFNNs).

# Introduction

Predicting the state of complex, nonlinear dynamical systems as a function of time is an important problem of great practical utility. Recent advances of artificial neural networks and machine learning (ML) methods have made possible significant applications in science, industry, and technology (1-14), with reliable prediction comprising one of the most promising areas of research. In this report, we investigate the long-term forecasting capability of two widely used ML methods, the Long Short-Term Memory (LSTM) and the reservoir-computing (RC) recurrent neural network architectures, to predict the spatiotemporal evolution of two distinct complex dynamical phenomena: (i) multi-clustered turbulent chimera states, that is collective, self-organized patterns of coexisting coherence and incoherence in coupled oscillators systems; and (ii) the onset of branching singularities in electronic flow in two dimensional disordered materials. These two phenomena represent opposite ends in complex spatiotemporal evolution with chimeras relating to dynamic self-organization and branching relating to the stochastic onset of singular motion.

LSTM networks (4) have proven to be successful in predicting sequence-involving tasks such as speech recognition (5), machine translation (6), and human dynamics (7) among others. While they have the ability to learn and reproduce long, isolated sequences, they do not seem to provide robust predictions in complex physical systems that involve chaotic behavior or to capture dependencies between multiple correlated sequences (8-13); in many cases they are outperformed by simpler methods, such as residual multilayer perceptrons (8), while in other cases their predictions diverge (9, 13). Recently, hybrid LSTM models have been proposed to improve forecasting in applications ranging from chaotic systems (8, 9), to pedestrian trajectories (10), particle tracking in high-energy physics (11), and anomaly detection in the post-mortem time series of superconducting magnets (12). For example, in the method of predicting human trajectories by "social LSTMs" (13), one LSTM network is assigned to each pedestrian, but in order to improve the prediction, each LSTM network is not independent of all other LSTMs, but are connected to those corresponding to nearby pedestrians' sequences by the introduction of a "social" pooling layer which allows the LSTMs of spatially proximal sequences to share their hidden-states with each other. In another approach, aiming at improving the forecasting of higher-dimensional chaotic systems, a hybrid architecture extends the global LSTM applied to the system with a mean stochastic model (MSM-LSTM) in order to ensure convergence to the invariant measure (9).

Reservoir computing (RC), first proposed by Jaeger and Haas (14), comprises a linear input layer, a recurrent nonlinear reservoir network and a linear output layer. This approach has recently been applied to inference problems in chaotic systems (1, 15, 16). Specifically, "observers" trained by RC (the reservoir observers), provide continual "ground truth" system state measurements as input to the prediction method, deducing the state of the chaotic dynamical system as a function of time from the limited number of the concurrent system state measurements. At each time step, RC estimates the desired unmeasured variables from the measured variables and predicts the evolution the physical system. Lu *et al.* (3) have recently demonstrated the effectiveness of the method by applying it to the Rossler system, the Lorenz system, and the spatiotemporally chaotic Kuramoto–Sivashinsky equation, carefully pointing out that the method addresses the inference of unmeasured state variables rather than their prediction.

**Introducing a new method: "Observer LSTM" (OLSTM)**
LSTM networks have proven successful in learning and generalizing sequential tasks from isolated sequences, such as handwriting and speech. Inspired by this success, we first considered a model with a single LSTM network assigned to each system node, which is independent of all other LSTMs. The prediction error for this approach turned out to be very large. This is not surprising since chimera states are collective phenomena and the simplistic use of one LSTM network per node, independent of all others, does not capture well the interaction between the nodes; the sequences of different nodes are not isolated but correlated.

In order to address the independent LSTMs' limited ability to capture dependencies between multiple correlated sequences, we propose and demonstrate a new method, which we call "Observer LSTM (OLSTM)," based on the extension of the notion of "reservoir observers" to the LSTM networks. In the OLSTM method, we assign one LSTM to each (non-observer) system node but also assign "observer" status to certain system nodes ("LSTM observers", taken at equidistant positions for simplicity) which provide continual "ground truth" measurements as input to the prediction method. We demonstrate that their presence even in small numbers (of order less than 10% of the total number of nodes) greatly improves the long-term data-driven forecasting capability of the LSTM networks and provides the framework for a consistent comparison between the RC and LSTM methods.

Time-series data are used to train each network while no knowledge of the underlying system equations is required. Each individual LSTM network is trained by taking as input a number of past values (denoted as $N_p$) for the node at hand, plus the ground-truth values provided by all observers. Thus, the OLSTM method generates a generalized sequence, which combines the $N_p$ past values and the time-varying systemic interaction. Using the trained networks, long-term predictions are made by iteratively predicting one step forward for each node, using as input the node's previous values and the values provided by the relatively very small number of observers.

We study the networks' long term forecasting capability as a function of the number of observers and as a function of training-set size, and we compare the OLSTM performance with that of "reservoir observers" trained by RC, which utilizes a single ("global") network for the entire system. We benchmark both methods against a standard Feed-Forward neural network (FNN) method, with the same number of observers (OFNN). We compare quantitatively the networks' performance by calculating the root mean square error (RMSE) at each time-step, for all system nodes, over the predicted time-steps, as in (3), Eq. (21), without counting observer nodes and training time-steps, since they do not contribute to the prediction error.

## Results & Discussion

We structure our study of predicting complex dynamics in extended physical systems with ML approaches in two parts: the first part concerns chimeras and the second concerns branching in flows. Together, these two cases capture extremes of complex spatiotemporal behavior that severely challenge any method aspiring to predict the long-term behavior. Chimera states challenge the ML methods to predict partial, self organized coherence while the stochastic onset of branching challenges the ML methods to predict stochastic yet singular events.

**A. Predicting turbulent chimeras in coupled arrays.**

Chimera states are collective, self-organized patterns of coexisting coherence and incoherence in coupled oscillator systems. Following the first discovery of chimeras for symmetrically coupled Kuramoto identical oscillators in 2002 (17), this counterintuitive symmetry breaking of partially coherent and partially incoherent behavior has received enormous attention. Many recent theoretical works have focused on the study of chimera states in a variety of physical systems such as superconducting metamaterials (18, 19, 20) quantum systems (21), and laser arrays (22, 23), to mention only a few. Chimeras have also been studied in models addressing neuron dynamics in hierarchical and modular networks (24, 25). It has been suggested that chimera states may be related the phenomenon of unihemispheric sleep observed in mammals and birds (26), epileptic seizures (27), and blackouts in power-grids (28). For finite systems, chimera states are known to be chaotic transients (29), can be stabilized by various newly developed control schemes (30, 31, 32). Following the theoretical predictions, chimera states were experimentally verified for the first time in populations of coupled chemical oscillators (33) and in optical coupled map lattices realized by liquid-crystal light modulators (34) and, later on, in mechanical (35), electronic (36), and electrochemical oscillators (37); for a recent review, see (38).

Chimeras can be stationary or turbulent. Turbulent chimeras have been observed experimentally (34) and have been classified in numerical studies of large arrays of SQUIDs (Superconducting QUantum Interference Devices) and in arrays of lasers with various types of interactions (18-20, 22), among other physical systems. Their actual trajectories are highly nonlinear and comprise an immense challenge to predicting their occurrence. In the following sections, we present the long-term prediction results of the ML methods applied on turbulent chimeras in simulated SQUID and semiconductor laser arrays.

SQUID metamaterials constitute a subclass of superconducting artificial media whose function relies both on the geometry and the extraordinary properties of superconductivity and the Josephson effect (39, 40). Recent experiments on SQUID metamaterials in the superconducting state have demonstrated their wide-band tuneability, significantly reduced losses, and dynamic multistability (41, 42). The simplest version of a SQUID consists of a superconducting ring interrupted by a Josephson junction (43); the device is a highly nonlinear resonator with a strong response to applied magnetic fields. SQUID metamaterials exhibit peculiar magnetic properties including negative diamagnetic permeability that were predicted both for the quantum (44) and the classical (45, 46) regime.

We investigate the long-term prediction capability of the ML methods under study on turbulent single-headed and double-headed chimeras ("head" stands for incoherent cluster) observed numerically in an array of N identical *rf* SQUIDs coupled together magnetically through dipole-dipole forces. In this system, the magnetic flux $\Phi_n$ threading the n-th SQUID loop is given by:

$$\Phi_n = \Phi_{ext} + LI_n + L \sum_{m \neq n} \lambda_{|m-n|} I_m,$$

where the indices *n* and *m* run from 1 to N, $\Phi_{ext}$ is the external flux in each SQUID, $\lambda_{|m-n|} = \frac{M_{|m-n|}}{L}$ is the dimensionless coupling coefficient between the SQUIDs at positions *m* and *n*, $M_{|m-n|}$ being their corresponding mutual inductance, and

$$-I_n = C \frac{d^2 \Phi_n}{dt^2} + \frac{1}{R} \frac{d\Phi_n}{dt} + I_c \sin\left(2\pi \frac{\Phi_n}{\Phi_0}\right)$$

is the current in the n-th SQUID given by the resistively and capacitively shunted junction (RCSJ) model (47), with $\Phi_0$ the flux quantum and $I_c$ the critical current of the Josephson junctions. Each individual SQUID is a highly nonlinear oscillator exhibiting multistability in a certain parameter regime. This is crucial for the occurrence of the chimera states when considering the coupled system. The number of possible states in a SQUID metamaterial is not merely the sum of the combinations of individual SQUID states, since their collective behavior provides many more possibilities. Depending on the choice of initial conditions, various space-time flux patterns may be obtained. "Wild" turbulent chimeras are important for testing the prediction methods because they offer non-trivial dynamical evolution (trajectories) upon which we can test the long-term data-driven (model-free) forecasting of the ML methods. Such chimeras have been generated in Hizanidis *et al.* (19) (see their Fig. 4d for a double-headed chimera with large size of incoherent clusters and Fig. 4e for a single-headed chimera, in which the largest part of the metamaterial is occupied by an incoherent cluster with varying size and position in time), where the evolution of the individual SQUID fluxes $\Phi_n$ is monitored.

We apply RC, OLSTM, and OFNN methods for long-term prediction of the dynamics of these single-headed and double-headed chimeras. Prediction snapshots for the single-headed chimera are presented in Fig. 1*A* as insets on the actual (ground-truth) spatiotemporal evolution of the fluxes. In predicting the evolution of this chimera, 17 "observers" were used in the positions marked by the tips of the arrows. The predicted time series for the flux of SQUID #223 is being shown (Fig. 1*A*, right vertical inset) as well as the predicted fluxes for all SQUIDs (entire metamaterial) at time-step $t_n$=11,000 (Fig. 1*A*, bottom inset). Similar results for the double-headed chimera are presented in Fig. 1*B*, also for 17 observers; the predicted time series for the flux of SQUID #163 is being shown in Fig. 1*B* (right vertical inset) along with the predicted fluxes for all SQUIDs at time-step $t_n$=12,000 (Fig. 1*B*, bottom inset). The specific SQUID nodes (#223 and #163) have been chosen in order to depict the long-term forecasting capability of the ML methods in challenging regimes that include *both* coherent and incoherent behavior. We emphasize that in both cases these are very long-time predictions, not just short-term prediction of just a few time steps beyond the training time; specifically, these predictions comprise more than twice (for OLSTMs and OFNNs) and three times (for RC) the training time. Nevertheless, the ML methods produce non-divergent predictions.

In our implementation of OLSTMs we found that a large value of the number of past steps $N_p$ is not necessary, and in fact even with $N_p = 1$ the results are quite satisfactory; this implies that the presence of observers more than compensates the need for a short memory to guide the predictions. Thus, our results demonstrate that the "neighborhood" of the node is important; it changes dynamically and the assistance of observers in predicting the future values is more appropriate to process the temporal variation of the "neighborhood" state. The same considerations apply to both SQUID-array and laser-array chimeras.

In laser systems, chimeras were first reported both theoretically and experimentally in a virtual space-time representation of a single laser system subject to long delayed feedback (33); small networks of globally delay-coupled lasers have also been studied and chimera states were found for both small and large delays. In the present report, we apply the ML methods to predict the trajectories of the turbulent chimeras presented in Fig 4V of the recent work of Shena *et al.* (22) on multi-clustered chimeras in a large array of coupled semiconductor lasers with nonlocal coupling. This array is a ring of $M = 200$ semiconductors lasers of class B. Each node $j$ is symmetrically coupled with the same strength to its $R$ neighbors on either side (nonlocal coupling). The evolution of the slowly varying complex amplitudes $\mathcal{E}_j = E_j \exp(i\varphi_j)$, with $E_j$ is the amplitude and $\varphi_j$ the phase of the electric field, and the corresponding population inversions $N_j$ are given by equations

$$\frac{d\mathcal{E}_j}{dt} = (1 + ia)\mathcal{E}_j N_j + \frac{ke^{-i2C_p}}{2R} \sum_{l=j-R}^{j+R} \mathcal{E}_l,$$

$$\frac{dN_j}{dt} = \frac{1}{T}[p - N_j - (1 + 2N_j)|\mathcal{E}_j|^2], \quad j = 1, \cdots, M,$$

where all indices have to be taken modulo $M$. $T$ is the ratio of the lifetime of the electrons in the excited level and that of the photons in the laser cavity. Lasers are pumped electrically with the excess pump rate $p = 0.23$ (22). The linewidth enhancement factor $a$ models the relation between the amplitude and the phase of the electrical field; a value of $a = 2.5$ was used, typical for semiconductor lasers. The coupling strength $k$ and the phase $C_p$ are the control parameters that are used to tune the collective dynamics of the system. As a measure for phase and amplitude synchronization, the characterization of the phase synchronization of the system can be calculated by means of the Kuramoto local order parameter (48):

$$Z_j = \left|\frac{1}{2\zeta} \sum_{|l-j|\leq\zeta} e^{i\varphi_l}\right|$$

Shena *et al.* [22] have used a spatial average with a window size of $\zeta = 3$ elements (a $Z_j$ value close to unity indicates that the $j$-th laser belongs to the coherent regime, whereas $Z_j$ is closer to 0 in the incoherent part). This quantity can measure only the phase coherence and gives no information about the amplitude synchronization of the electric field, and for this reason the authors have used the classification scheme presented in (49) for spatial coherence, based on the calculation of the so called *local curvature* at each time instance, by applying the absolute value of the discrete Laplacian |*DE*| on the spatial data of the amplitude of the electric field:

$$|DE|_j(t) = |E_{j+1}(t) - 2E_j(t) + E_{j-1}(t)|, \quad j = 1, \cdots, M.$$

In the synchronization regime the local curvature is close to zero while in the asynchronous regime it is finite and fluctuating.

Since our objective is to obtain "wild" turbulent chimeras in order to test the long-term predictions of the ML methods, we have chosen the turbulent chimera of (22) Fig 4V, generated with the choice of parameter values: $R = 64$, $C_p = 0.4\pi$, $k = 0.225$, $T = 392$, $p = 0.23$, $a = 2.5$. Prediction snapshots are presented in Fig. 2, as insets on the actual (ground-truth) spatiotemporal evolution of the local curvature for the flux of laser #145 and for all lasers, for similar choices as in Fig. 1, that is, with 17 "observers" and at time-step $t_n=1,500$. These are extremely long-time predictions, with time horizons almost five times the training time (for OLSTMs and OFNNs) and ten times (for RC) the training time. In this case, as in that of the SQUID system, all ML methods under study produce non-divergent predictions.

In Fig. 3 we present the calculated values of the mean squared error (RMSE) for long-term prediction (top inset) of the RC, OLSTM, and the OFNN methods, and as a function of the number of observers and the size of the training dataset for a randomly selected time-step for the chimera depicted in Fig. 2; similar results are obtained for all systems. The RMSE values remain, on average, close to $10^{-1}$, for time horizons up to 2,000 time-steps in the future. This figure also presents graphs of how the RMSE varies as a function of the number of observers (given as percentage of the number of the system nodes), and as a function of the size of the training datasets (given as percentage of the entire ground-truth time series. In all cases, low RMSE values are attained after a minimum number of observers have been included in the system, comprising about 5% of the total number of oscillators, and the size of the training data reaches at least 15% of entire dataset. These results demonstrate that, even very small numbers of observers are very important for achieving non-diverging long-term predictions. Furthermore, they show that RC achieves low levels of RMSE with smaller size of training datasets, whereas OLSTM achieves low levels of RMSE with smaller number of observers.

**B. Predicting singular events**

Wave focusing due to refractive index variation is a common occurrence in many physical systems, as, for example, in optical media where the index of refraction changes in a statistical way due to small imperfections or distributions of defects in the medium through which the wave propagates. Random spatial variability of the index leads to local focusing and defocusing of the waves and the formation of caustics (or wave "branches") with substantially increased local wave intensity (50, 51). Quantum particles like electrons also exhibit wave properties and as a result, electrons travelling in disordered media can form coalescing trajectories and exhibit phenomena similar to wave motion in optical random media. A case in point, where branching may have implications for technological applications, is the ultra-relativistic electronic flow in a two-dimensional (2D) random potential as manifested in graphene and other Dirac solids (51). In these situations, branching arises in the flow of electrons through a region of inhomogeneous distribution of charge impurities in the substrate, which create a random potential for the electrons (52). An additional bias voltage is introduced to induce the electronic propagation. Recently, it was shown that the onset of the electronic branched flow is determined by the statistical properties of the random substrate potential, which is quantified through a scaling-type relationship to capture the emergence of branches (53).

It is a formidable challenge to predict singular events like branching in wave propagation or electron flow because of the stochastic nature of the onset of such events. An important question is whether or not ML methods can dissect the stochastic nature of branching and "learn" the interactions that take place among trajectories, thereby providing an accurate detection mechanism for the caustics that mark the onset of branching. We attempt to resolve this issue with results from the RC, OLSTM and OFNN methods, on singular branched flows in graphene with random potentials. Our results demonstrate that the ML methods we considered can adequately capture the stochastic temporal dependencies of the time series in this prototypical complex dynamical system.

Caustic event prediction in a 2D electron flow is facilitated if we consider one of the spatial dimensions (we refer to it as the "longitudinal" x-direction) as the "time-coordinate", and therefore, map the stationary phenomenon of caustic formation onto a 1D spatio-temporal dynamical problem. In the framework of this approach, we model the motion of electrons as individual rays whose density matrix is transformed onto a vector of time series with dimension $N$, the number of mesh points of the remaining spatial dimension (the "transverse", or y-direction). In the system we studied, $N=210$, spanning a total length of 84 nm (53). Our goal is to predict the onset of branching in time and its location in the electronic flow. Fig. 4 presents prediction snapshots of the onset of branching in electronic flows in graphene with random potentials, depicting the intensity of the flows, using 10 different "observers" along the "time" axis of the flows at different positions on the transverse axis. These results demonstrate that the RC, OLSTM and OFNN methods are able to predict well the branching in the electronic flows in this system, even at very long prediction times. In these methods, observers are very important to achieve long-term forecasting capability; they act as real-time sensors that help predict the future values of the flows.

## Conclusions and Outlook

The issue of predicting complex spatiotemporal behavior using ML approaches is one of central importance for their potential applications in the physical sciences and beyond. Here, we attempted to address this issue by considering two distinct prototypical phenomena, viz. partially coherent chimera states and the stochastic onset of branching in 2D wave flows, as these are realized in coupled arrays of SQUIDs or lasers, and in the flow of electrons in graphene with random potentials, respectively. We find that ML approaches like LSTM and RC recurrent neural networks can perform well in predicting complex dynamics in extended physical systems when they involve "observers" that monitor the system evolution throughout its time dimension. The presence of observers is an inherent requirement of the second approach (RC), but not of the first (LSTM). Accordingly, we proposed a new method, which we call "Observer LSTM (OLSTM)," to address the limitations of single, independent LSTM networks in capturing dependencies between multiple correlated sequences. We have also considered an observer-enhanced Feed-Forward network (OFNN) and tested the long-term prediction performance of the three approaches, OLSTM, OFNN, and reservoir observers trained by RC, on the two difficult problems of turbulent chimeras and 2D branching flows. Our results quantify how the prediction error (root mean squared error, RMSE, of the predicted values) varies as a function of the number of observers and of the size of the training datasets.

We conclude that "observers" comprise *sine qua non* elements for robust data-driven (model-free) long-term forecasting capability of the LSTM and FNN networks, and they provide the framework for a consistent comparison between the LSTM and RC methods. Thus, observer-enhanced ML methods, like OLSTM, acting as high-level "intelligent" interpolation schemes, are capable of successfully predicting the nonlinear spatiotemporal evolution of complex dynamical systems. Many issues remain to be evaluated and questions to be answered in establishing the robustness and efficiency of the observer-enhanced approaches. An example of such issues is the extent to which the presence of observers, while enhancing long-term predictability, affects the chaotic behavior of the system. A thorough investigation of such issues is the focus of on-going and future research.

# Methods

The RC network architecture used in predicting turbulent chimeras comprises 1000 reservoir nodes, with spectral radius $\rho = 1.0$, average degree $D = 80$, scale of inputs weights $\sigma = 1.5$, bias constant $\xi = 0.0$, leakage rate $\alpha = 0.9$, ridge regression parameter $\beta = 0.5$, and time interval $\Delta t = 0.01$. The RC network applies to the system as a whole (single network architecture).

The RC network architecture used in predicting the onset of branching comprises 3000 reservoir nodes, with spectral radius $\rho = 0.9$, average degree $D = 50$, scale of inputs weights $\sigma = 1$, bias constant $\xi = -0.4$, leakage rate $\alpha = 0.5$, ridge regression parameter $\beta = 0.05$, and $\Delta t = 0.05$. The RC network applies to the system as a whole (single network architecture).

The OLSTM network architecture used in this study comprises 400 LSTM cells with RELU activation functions (one hidden layer). A single LSTM network is applied to each (non-observer) system node (SQUID or laser oscillator or electron flow). Similarly, a single fully-connected (dense) Feed Forward network (OFNN), with RELU activation functions, is applied to each (non-observer) system node (SQUID or laser oscillator or electron flow).

The data of the time series used (chimeras and electronic flows in graphene) was preprocessed as in (3) Eq. (14) and smoothed by means of Matlab's 2D Gaussian Smoothing Kernel with standard deviation $\sigma=3.5$ for chimeras and $\sigma=2.5$ for the electronic flows in graphene.

The time series data have been divided into two separate sets, the training dataset and the validation dataset. The data is stacked in batches (of size 50 data points) in order to form the training (and validation) input and output of the networks. These training batches are used to optimize the parameters of the networks (weights and biases). The training proceeds by optimizing the network weights iteratively for each batch (training of one epoch). The training loss function is a weighted version of the root mean square error.

Optimization for LSTMs during training is performed using the Adam stochastic optimization method (54) with a learning rate of 0.001 as implemented in *Keras* (55).

Training is stopped after 200 epochs and the OLSTM (OFNN) network with the smallest validation error is selected in order to avoid over-fitting. Each trained LSTM network is then used to forecast the node's state in the next time step in an iterative fashion (getting also input from the observers).

As a comparison measure for the networks' performance, we use the root mean square error (RMSE) calculated at each time-step, for all system nodes and over the predicted time-steps (unless otherwise specified).

**ACKNOWLEDGEMENTS**

G.P.T. acknowledges useful discussions with Prof. Edward Ott. G.N. and G.P.T. acknowledge support by the European Commission under project NHQWAVE (MSCA-RISE 691209). J.H. acknowledges support by the Greek State's General Secretariat for Research and Technology (GSRT) via the Postdoctoral Researchers Projects Fund. M.M. and E.K. acknowledge partial support from NSF grant 1542807, EFRI 2-DARE: Quantum Optoelectronics, Magnetoelectronics and Plasmonics in 2Dimensional Materials Heterostructures.

**Authors' Contributions**: G.N., M.M., and G.D.B. designed and performed research, analyzed data, and wrote the manuscript; J.H. provided data on chimeras, made contributions to the manuscript, and constructive comments; G.P.T. and E.K. contributed to the design of research, interpretation of results and to the writing of the manuscript.

**Data Accessibility Statement**

The machine learning library *Keras* in python 3.6 was used for the implementation of the OLSTM and OFNN architectures (utilizing the "Metropolis" Supercomputing Facility of the Center for Quantum Complexity and Nanotechnology of the Physics Department of the University of Crete) while Matlab was used for the RC architectures. Data and code used to produce the results will be available online upon publication.

**Competing Interests Statement**

We have no competing interests.

**FIGURES & FIGURE CAPTIONS**

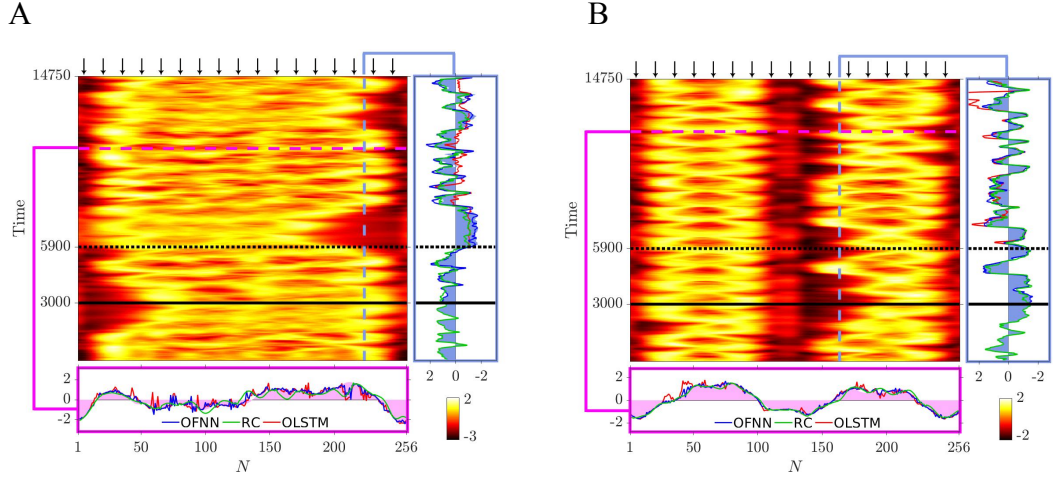

**Figure 1**. Spatiotemporal plots of single- and double-headed chimeras and predicted time series and fluxes. *(A)* Spatiotemporal plot of a single-headed chimera state, generated in the 1-dimensional SQUID array of Ref (19) Fig. 4e, depicting the evolution of the fluxes (values are color-coded), for the one-headed chimera observed in a large array of 256 coupled SQUIDs by dipole-dipole moments, which has been studied numerically for a nonlocal coupling scheme. In predicting the spatiotemporal evolution of this chimera, 17 "observers" have been placed in the positions marked by the tips of the arrows. The thick horizontal black line marks the end of the RC training time, while the dotted horizontal black line marks the end of the OLSTM (and OFNN) training time. Right, vertical inset: predicted time series for the flux of SQUID #223: shaded blue color depicts the actual (ground-truth) data, red line depicts prediction by OLSTM, green line depicts prediction by RC, blue line depicts prediction by OFNN. Bottom inset: predicted fluxes for all SQUIDs (entire metamaterial) at time-step $t_n=11,000$: symbols are the same as in the right inset, but with pink depicting the actual (ground-truth) data. *(B)* same as in part a), but for a double-headed chimera state generated in the 1-dimensional linear SQUID array of Ref (19) Fig. 4d. Right, vertical inset: the predicted time series for the flux of SQUID #163. Bottom inset: the predicted fluxes for all SQUIDs (entire metamaterial) at time-step $t_n=12,000$.

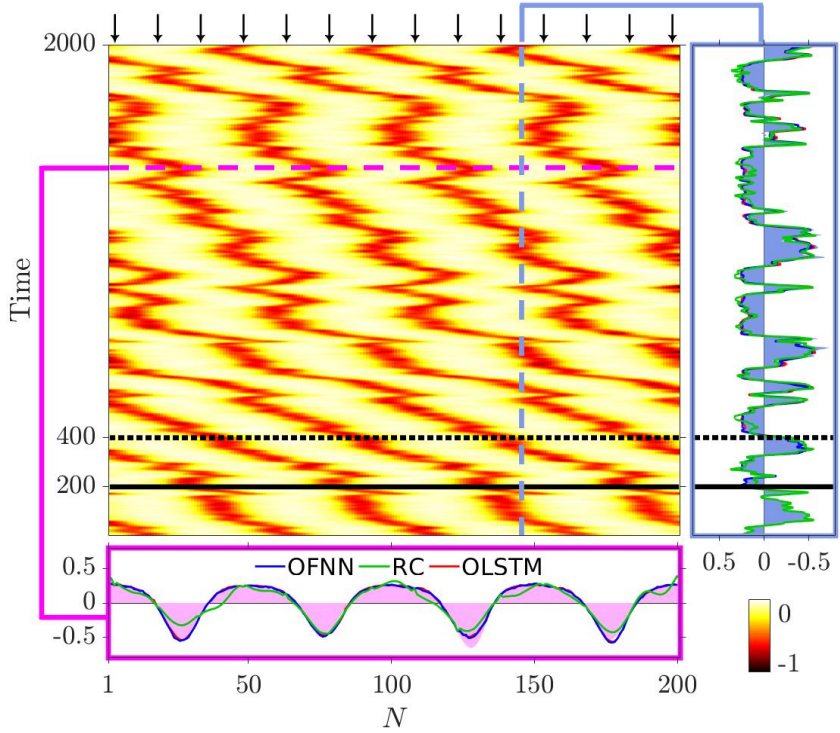

**Figure 2**. Spatiotemporal plot of a turbulent chimera state (as is generated in the 1-dimensional semiconductor Class B laser array of (22) Fig. 4V), depicting the evolution of the local curvature for each laser (values are color-coded), studied numerically for a nonlocal coupling scheme. 17 "observers" have been placed in the positions marked by the tips of the arrows. The thick horizontal black line represents the size of the training dataset used for training the reservoir observers with RC, while the dotted horizontal black line represents the size of the training dataset for training the OLSTM (and OFNN) observers. Right, vertical inset: the predicted time series for laser #145: shaded blue color depicts the actual (ground-truth) data, red line depicts prediction by OLSTM, green line depicts prediction by RC, blue line depicts prediction by OFNN. Bottom inset: snapshot of the spatial profile of the predicted local curvatures for the entire array at time-step $t_n=1,500$: color code is as in the right inset, but with pink color depicting the actual (ground-truth) data.

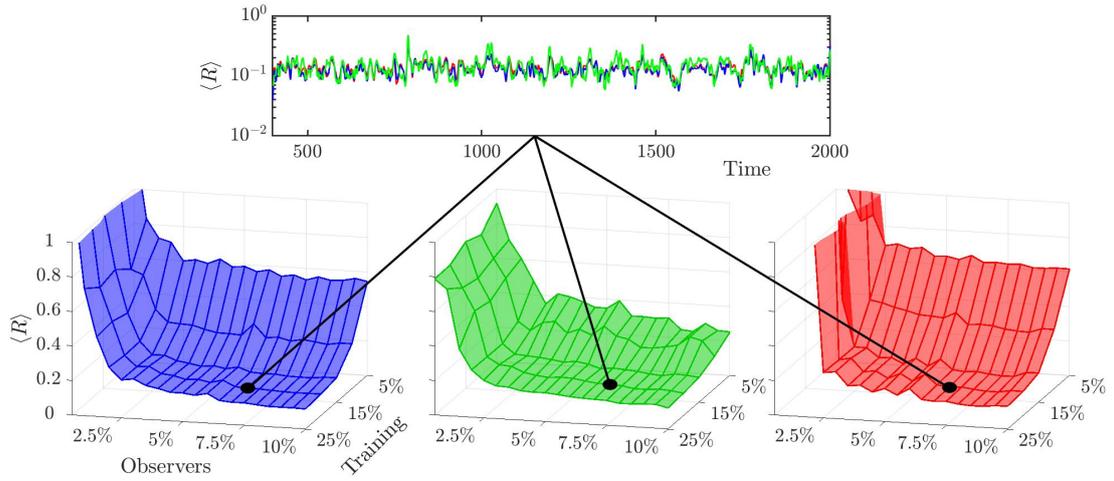

**Figure 3**. RMSE (<R>) of RC, OLSTM, and the OFNN methods calculated for each time-step over all system nodes at each predicted time-step (top panel) and over the predicted time-steps *and* all system nodes (bottom 3-dimensional plots) as a function of the number of observers (given as percentage of the entire number of system nodes) and the size of the training dataset (given as percentage of the entire ground-truth time series) for the chimera depicted in Fig. 2. The black dots in each of the 3-dimensional plots represent the RMSE for each ML method, respectively, at a randomly picked time-step, calculated over all system nodes *and* previous time-steps, but not counting observer nodes and training time-steps. Red: OLSTM, green: RC, blue: OFNN. Axes, tick marks, tick labels, and grid lines are the same in all 3-dimensional plots.

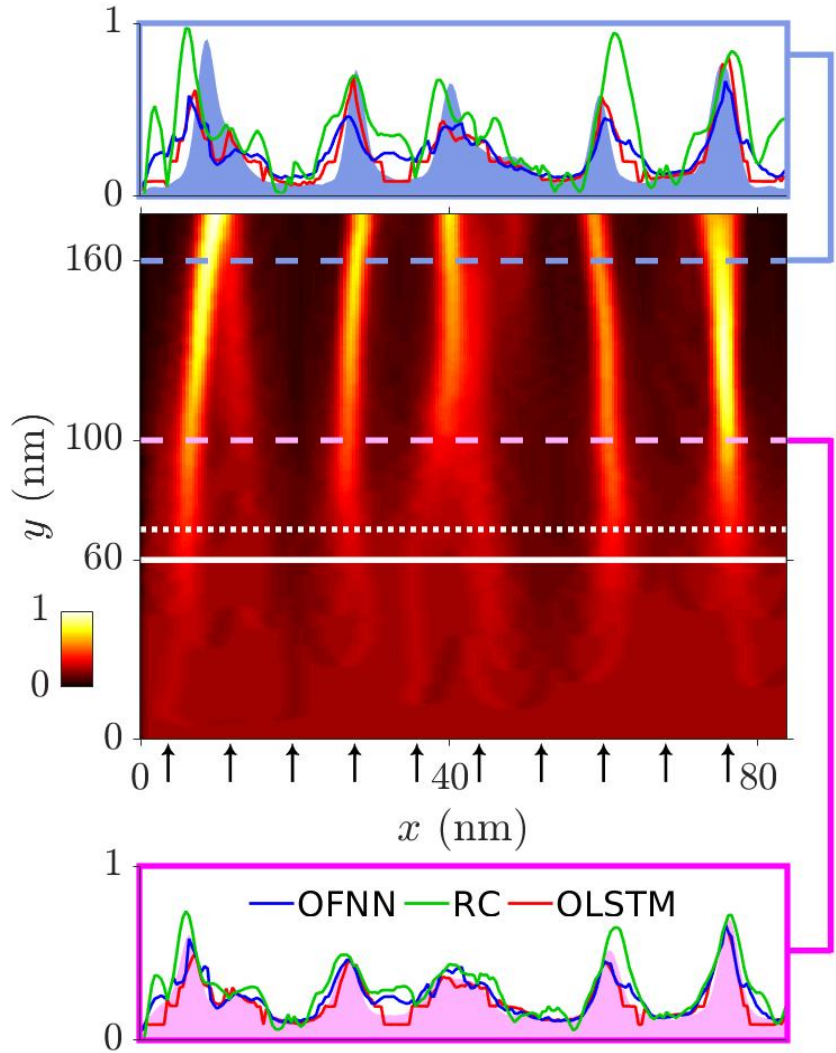

**Figure 4**. Branching in electronic flows in graphene with random potentials (intensity values are color-coded); the graphene sheet has size 176 nm (vertical) x 84 nm (horizontal). The snapshots in insets depict the intensity of the flows as found in the simulation and predicted by the ML methods. In predicting the time evolution of the flows, 10 "observers" have been placed in the positions marked by the tips of the arrows, monitoring the entire "time" (vertical) axis. The thick horizontal white line marks the end of the RC training, while the dotted white line marks the end of the OLSTM (and OFNN) training. Inset outlined in pink: the actual and predicted time series for the entire system at "time coordinate" point $x_n$=100 nm, corresponding to 501 time-steps. The pink-shaded curve represents the actual (ground-truth) data, red line is the OLSTM prediction, green is the RC prediction and blue the OFNN prediction. Inset outlined in blue: same as for the other inset, but at "time coordinate" point $x_n$=160 nm, corresponding to 801 time steps, and with blue-shaded curve depicting the actual (ground-truth) data.